\documentstyle[preprint,seceq,mbf]{ptptex}
\notypesetlogo 
 \def\ind{\indent}
 \def\nn{\nonumber}

 \def\be{\begin{equation}}
 \def\ee{\end{equation}}
 \def\ben{\begin{enumerate}}
 \def\een{\end{enumerate}}
 \def\bl{\begin{flushleft}}
 \def\el{\end{flushleft}}
 \def\bt{\begin{tabular}}
 \def\et{\end{tabular}}
 
 \def\br{\begin{flushright}}
 \def\er{\end{flushright}}
 \def\bc{\begin{center}}
 \def\ec{\end{center}}
 \def\bea{\begin{eqnarray}}
 \def\eea{\end{eqnarray}}
 \def\bea*{\begin{eqnarray*}}
 \def\eea*{\end{eqnarray*}}
 \def\ba{\begin{array}}
 \def\ea{\end{array}}
 \def\bi{\begin{itemize}}
 \def\ei{\end{itemize}}

 \def\cL{{\mbox {${\cal L}$}}}
 
 \def\cA{{\mbox {${\cal A}$}}}
 
 \def\cM{{\mbox {${\cal M}$}}}
 
 \def\cU{{\mbox {${\cal U}$}}}
 \def\bA{{\mbox {{\mbf A}}}}

 \def\bC{{\mbox {{\mbf C}}}}
 
 \def\bF{{\mbox {{\mbf F}}}}
 \def\bH{{\mbox {{\mbf H}}}}
 \def\bG{{\mbox {{\mbf G}}}}

 \def\bd{{\mbox {{\mbf d}}}}

 \def\tbA{{\mbox {${\tilde {\!\!{\mbf A}}}$}}}

 \def\cham{Chamseddine}
 \def\ind{\indent}
\def\NP{Nucl. Phys.$\;$} 

\def\PL{Phys. Lett.$\;$}

\def\be{\begin{equation}}

 \def\cL{{\mbox {${\cal L}$}}}
 \def\cA{{\mbox {${\cal A}$}}}
 \def\cU{{\mbox {${\cal U}$}}}

 \def\cU{{\mbox {${\cal U}$}}}
 \def\bA{{\mbox {{\mbf A}}}}

 \def\bC{{\mbox {{\mbf C}}}}
 \def\bF{{\mbox {{\mbf F}}}}
 \def\bH{{\mbox {{\mbf H}}}}
 \def\bG{{\mbox {{\mbf G}}}}
 
 \def\bd{{\mbox {{\mbf d}}}}

 \def\dslash{\partial{\raise 1pt\hbox{$\!\!\!/$}}}
 \def\dslash{\partial{\raise 1pt\hbox{$\!\!\!/$}}}
  \newfont{\bg}{cmr10 scaled\magstep4}
 \newcommand{\bigzerol}{\smash{\lower0.7ex\hbox{\bg 0}}}
 \newcommand{\bigzerou}{%
    \smash{\lower0.5ex\hbox{\bg 0}}}
 \newcommand{\bigzeroll}{\smash{\lower1.7ex\hbox{\bg 0}}}
 \newcommand{\bigzerouu}{%
    \smash{\lower1.7ex\hbox{\bg 0}}}
 \newcommand{\bigzerolll}{\smash{\lower2.5ex\hbox{\bg 0}}}
 \newcommand{\bigzerouuu}{%
    \smash{\lower2.5ex\hbox{\bg 0}}}
 \newcommand{\bigzeroL}{\smash{\lower0.01ex\hbox{\bg 0}}}
\title{%
A Field-Theoretic Approach to Connes' Gauge Theory 
on $M_4\times Z_2$
}
\vspace{5mm}
\author{%
  Hiromi {\sc Kase}$^{1}$, Katsusada {\sc Morita}$^{2}$ and 
  Yoshitaka {\sc Okumura}$^{3}$
}
\inst{%
{\it  $^{1}$Department of Physics, 
  Daido Institute of Technology, Nagoya 457-0811, Japan\\
  $^{2}$Department of Physics, 
  Nagoya University, Nagoya 464-8602, Japan\\
  $^{3}$Department of Natural Sciences, 
  Chubu University, Kasugai, 487-0027, Japan}
}
\vspace{15mm}
\abst{%
Connes' gauge theory on $M_4\times Z_2$
is reformulated
in the Lagrangian level.
It is pointed out that
the field strength in Connes' gauge
theory is not unique.
We explicitly construct
a field strength
different from Connes' one and prove that
our definition
leads to the generation-number independent
Higgs potential.
It is also shown that
the nonuniqueness is related to
the assumption that
two different extensions
of
the differential geometry
are possible
when the extra one-form basis
$\chi$
is introduced to define the
differential geometry on $M_4\times Z_2$.
Our reformulation is applied to the standard model
based on Connes' color-flavor algebra.
A connection between the unimodularity condition
and the electric charge quantization
is then discussed in the presence or absence of $\nu_R$.
}
\begin{document}
\maketitle
\section{Introduction}
Connes' interpretation of the standard model
in non-commutative geometry (NCG)\cite{1),2),3)}
is based on the assumption that
{\it an algebra underlies the gauge symmetry}.
This assumption armed with the mathematical apparatus of NCG
allows us to define the Yang-Mills (YM) gauge theory
on general manifold, either continuous or discrete.
It is remarkable that
the spontaneously broken gauge theory
observed in Nature
belongs to Connes' YM on 
a discrete manifold. Thus 
the non-commutative
one-form on
a two-sheeted
Minkowski space-time $M_4\times Z_2$
combines\cite{1),2),3),4),5),6)} the YM gauge fields with the Higgs one
in the standard model and determines dynamics in the bosonic sector
through Connes' field strength $\bG$.
\\
\ind
In view of the mathematical niceties involved
it is important to extract a physical
information as simple as possible.
We, therefore, feel it
worthwhile looking for a more accessible way
of reformulating Connes' YM 
without the axioms of NCG, which
would disclose important features in the theory
from physical point of view.
In this paper we continue our previous work\cite{7)}
to derive the non-commutative
one-form on $M_4\times Z_2$
in the Lagrangian formulation.
\\
\ind
During our investigation
we find that
the field strength in Connes' YM is not unique.
The nonuniqueness is totally unrelated with Connes' ambiguity
problem\cite{1),2)} but rather
originates from the different associative products of the Dirac
matrices.
We explicitly construct a different field strength
$\bF$ than Connes' one $\bG$
by introducing
a new associative product
of the Dirac matrices including $\gamma_5$.
The new field strength
leads to the
generation-number independent Higgs potential,
while the quartic coupling constant derived from $\bG$
depends on the generation number.
We shall show a close connection of 
the nonuniqueness with the extended differential 
formalism\cite{8),9),10)}.
\\
\ind
The plan of this paper is as follows.
In the next section
we present
a field-theoretic approach to
Connes' YM
by taking into account of two different field strengths.
In \S 3 
we review two possible extensions\cite{9),10)}
of the ordinary differential geometry
and show that
they precisely lead to the two different
field strengths.
We shall derive in \S4 Asquish's
representation\cite{11)} of Connes' color-flavor algebra\cite{3)}
of the standard model
using the double sum prescription\cite{12)}
and discuss its consequence regarding 
the electric charge quantization
in the presence or absence of $\nu_R$.
The final section is devoted
to discussion.
\section{A field-theoretic approach to Connes' YM on $M_4\times Z_2$}
Suppose that an algebra $\cA$
underlies the gauge symmetry.
To explain what this means in our methodology
we remark that,
although an arbitrary element of the algebra $\cA$
never defines the symmetry transformation,
it is possible to regard the Hilbert space of spinors $\psi$
as an $\cA$-module such that
the gauge group is given by the unitary group,
$G=\cU(\cA)=\{g\in\cA;gg^{\dag}=g^{\dag}g=1\}$.
To meet this condition
$\cA$ must be a local, unital and involutive algebra.
We are then tempted to consider the local non-symmetry transformations
\be
\left\{
     \ba{l}
     \psi\to\rho(b_i)\psi,\\
     {\bar\psi}\to{\bar\psi}\rho(a_i),\\
     \ea
     \right.\;\;a_i,b_i\in\cA,
\label{eqn:2-1}
\ee
where $\rho$ is the
$\ast$-preserving representation of the algebra $\cA$.
The linearity of the algebra
fits\cite{1),2),3)} the concept of generation
in a neat way.
Along with the transformations (\ref{eqn:2-1})
we
take the sum over the index $i$ in the
Lagrangian level provided that
\be
\displaystyle{\sum_i}\rho(a_i)\rho(b_i)=1(\equiv 1_{{\rm dim}\rho}),
\label{eqn:2-2}
\ee
which expresses
the unity decomposition
\footnote{
Recall that
the unity has an infinite variety of decompositions.
For instance, the unit matrix
{\scriptsize$\left(
 \ba{cc}
 1&0\\
 0&1\\
 \ea
 \right)$}
equals {\scriptsize$\left(
 \ba{cc}
 \alpha^{*}&-\beta\\
 \beta^{*}&\alpha\\
 \ea
 \right)\left(
 \ba{cc}
 \alpha&\beta\\
 -\beta^{*}&\alpha^{*}\\
 \ea
 \right)$} with $|\alpha|^2+|\beta|^2=1$
or the sum of terms
 {\scriptsize$\left(
 \ba{cc}
 \gamma&0\\
 0&\gamma^{*}\\
 \ea
 \right)
 \left(
 \ba{cc}
 0&\delta\\
 -\delta^{*}&0\\
 \ea
 \right)+
 \left(
 \ba{cc}
 0&\delta\\
 -\delta^{*}&0\\
 \ea
 \right)
 \left(
 \ba{cc}
 -\gamma^{*}&0\\
 0&-\gamma\\
 \ea
 \right)
 +\left(
 \ba{cc}
 c&-ic\\
 -ic&c\\
 \ea
 \right)\left(
 \ba{cc}
 d&id\\
 id&d\\
 \ea
 \right)$} (for real $c,d$ with $cd=1/2$)
and so on.
The first form defines $SU(2)$,
whereas the second sum contains non-unimodular
matrices.
In fact, all matrices in the second sum belongs to
the algebra $\bH$ of the real quaternions.} and
leaves ${\bar\psi}\psi$ invariant,
so that we end up with the covariant derivative
$D_0+A$
with the YM gauge field
\footnote{
Connes' original definition
$A=\sum_ia_i[D_0,b_i]$ has nothing to do with
the transformations (\ref{eqn:2-1}).
In our interpretation which may also be regarded as
a mnemonic one without NCG mathematics,
$a_i$ and $b_i$ are only the transformation parameters,
not treated as the canonical variables,
but the connection
$A=\sum_ia_i[D_0,b_i]$ is assumed to be a
field variable, as in Connes' YM,
which is promoted to be a quantum field\cite{13)}.}
\be
A=\displaystyle{\sum_i}\rho(a_i)[D_0,\rho(b_i)]
\equiv \displaystyle{\sum_i}a_i[D_0,b_i],\;
D_0=i\dslash
\otimes 1_{{\rm dim}\rho},\;\gamma^0A^{\dag}\gamma^0=A,
\label{eqn:2-3}
\ee
where $1_n$ denotes $n$-dimensional unit matrix
and dim$\rho$ is the dimension of $\rho$.
Here and hereafter we omit the notation
$\rho$ for simplicity unless necessary.
\\
\ind
Since the non-commutative one-form (\ref{eqn:2-3})
depends on the Dirac matrices,
there must be an ambiguity 
in extracting the curvature
to be identified with the YM field strength.
We compare it with
Connes' ambiguity in defining the field strength
based on the sum (\ref{eqn:2-3}).
The latter ambiguity arises from the fact that
the exterior derivative $dA$ in Connes' field strength
\be
G=dA+A^2,\;dA\equiv \displaystyle{\sum_i}[D_0,a_i][D_0,b_i]
\label{eqn:2-4}
\ee
may not vanish even for $A=\sum_ia_i[D_0,b_i]=0$.
To see this note that
$
G=F-1_4\otimes X,
$
where 
\be
F=d\wedge A+A\wedge A,\;d\wedge A\equiv \displaystyle{\sum_i}[D_0,a_i]
\wedge [D_0,b_i]
\label{eqn:2-5}
\ee
is the YM field strength
$F=(D_0+A)\wedge(D_0+A)=-(1/4)[\gamma^\mu,\gamma^\nu]F_{\mu\nu}$ with
the wedge product
of the Dirac matrices
\be
\gamma^\mu\wedge\gamma^\nu=\displaystyle{{1\over 2}}
(\gamma^\mu\gamma^\nu-\gamma^\nu\gamma^\mu)=\displaystyle{{1\over 
2}}[\gamma^\mu,\gamma^\nu],
\label{eqn:2-6}
\ee
and $X=C+A_\mu A^\mu$
with $C=\sum_i\partial_\mu a_i\partial^\mu b_i$.
Consequently, $G|_{A=\sum_ia_i[D_0,b_i]=0}=-1_4\otimes C$.
(One may replace $C$ with $C'=-\sum_ia_i\partial^2b_i$.)
This implies\cite{1),2)}
that the field strength in Connes' YM is to be defined as an equivalence
class, $[G]=[G']$ if $G=G'+\sum_i[D_0,a_i][D_0,b_i]$ with
$\sum_ia_i[D_0,b_i]=0, a_i,b_i\in\cA$.
Thus $[G]=F$ because the subtracted piece must be covariant.
In other words, if we define the field strength in Connes' YM
using the wedge product (\ref{eqn:2-6}),
we directly obtain the YM field strength.
In this sense Connes' ambiguity is related to the
ambiguity alluded to above.
As we shall see later,
this is no longer the case if Higgs is generated.
\\
\ind
There is an alternative method\cite{5)} to achieve the result $[G]=F$.
Although $X$ is gauge-covariant,
$C$ is not covariant and has no kinetic energy term 
in
the bosonic Lagrangian defined by the trace of the square of $G$
\be
\cL_B=-\displaystyle{{1\over 8}}{\rm Tr}
\displaystyle{{1\over g^2}}G^2=
-\displaystyle{{1\over 8}}{\rm Tr}
\displaystyle{{1\over g^2}}F^2-\displaystyle{{1\over 2}}
{\rm tr}\displaystyle{{1\over g^2}}X^2,
\label{eqn:2-7}
\ee
where
Tr also includes
the
trace over
Dirac matrices.
If we vary $A_\mu$ and $C$ independently,
we can eliminate
the auxiliary field $C$ through its 
equation of motion $X=0$.
Then the bosonic Lagrangian
(\ref{eqn:2-7}) is reduced to
the YM one
\be
\cL_{\rm YM}=-\displaystyle{{1\over 8}}{\rm Tr}
\displaystyle{{1\over g^2}}F^2=-\displaystyle{{1\over 8}}{\rm Tr}
\displaystyle{{1\over g^2}}[G]^2.
\label{eqn:2-8}
\ee
\ind
If the fermion mass matrix
$M$ is not gauge-invariant and
fermions exist in chiral multiplets,
we use the chiral decomposition of the spinors
so that the free Dirac operator reads
\be
D=D_0+i\gamma_5M,\;
D_0=\left(
    \ba{cc}
    i\dslash\otimes 1_{n_L}&0\\
    0&i\dslash\otimes 1_{n_R}\\
    \ea
    \right)\otimes 1_{N_g},\;
M=\left(
  \ba{cc}
  0&M_1\\
  M_1^{\dag}&0\\
  \ea
  \right).
\label{eqn:2-9}
\ee 
We then
obtain the gauge-invariant Dirac Lagrangian
$
\cL_{\mbf D}=\displaystyle{\sum_i}({\bar\psi}
\rho(a_i))D(\rho(b_i)\psi)
={\bar\psi}(D+\bA)\psi
$
with $\gamma^0\bA^{\dag}\gamma^0=\bA$,
where use has been made of the condition
(\ref{eqn:2-2})
and
\begin{eqnarray}
\bA&=&\displaystyle{\sum_i}a_i[D,b_i]
\label{eqn:2-10}\\[2mm]
&=&A+i\gamma_5\Phi,\;
A=\displaystyle{\sum_i}a_i[D_0,b_i]=i\gamma^\mu A_\mu,\;\;\;A_\mu=
\displaystyle{\sum_i}a_i\partial_\mu b_i,\;
\Phi=\displaystyle{\sum_i}a_i[M,b_i].
\nn
\end{eqnarray}
\ind
To define the curvature from the non-commutative one-form (\ref{eqn:2-10})
there again occur two kinds of ambiguity,
one intrinsic in Connes' YM and
the other coming from the different multiplication rules
of the Dirac matrices containing $\gamma_5$.
The first ambiguity is well-known.
We shall argue in the next section that
there are only two possible definitions.
In this section
we first consider
Connes'
field strength and then
our new
field strength which is obtained by generalizing the wedge product
(\ref{eqn:2-6}) to include $\gamma_5$.
\\
\ind
Connes' 
field strength
is given by\cite{1),2),3),4),5),6)}
\be
\bG=d\bA+\bA^2,\;d\bA\equiv \displaystyle{\sum_i}[D,a_i][D,b_i].
\label{eqn:2-11}
\ee
As noted before $d\bA$ may not vanish
even when $\bA=0$.
This makes it necessary to subtract a matrix $\langle\bG\rangle$
from $\bG$ where $\langle\bG\rangle$ is the matrix which is of
the {\it same form} as $d\bA$ with the constraint 
$\bA=0$
\footnote{As before the field strength
is defined as
an equivalence class, the equivalence
being given by
$\bG\sim \bG'$
if
$\bG'=\bG+\sum_i[D,a_i][D,b_i]$
with
$\sum_ia_i[D,b_i]=0,\;a_i, b_i\in \cA$.}.
We have seen above that, for $A$,
this is equivalent to putting $X=0$ in Eq.$\,$(\ref{eqn:2-7})
to obtain Eq.$\,$(\ref{eqn:2-8}) by the
variational principle.
The computation involved is not so simple for $\bA$.
A detailed mathematical account was given in Refs. 2), 4) and 6).
\\
\ind
The gauge-invariant bosonic Lagrangian
is
\begin{eqnarray}
\cL_B&=&-\displaystyle{{1\over 8}}{\rm Tr}
\displaystyle{{1\over g^2}}\bG^2
=\cL_{YM}+
\displaystyle{{1\over 2}}{\rm tr}
\displaystyle{{1\over g^2}}(D^\mu H)(D_\mu H)
-\displaystyle{{1\over 2}}{\rm tr}
\displaystyle{{1\over g^2}}Y^2,\nn\\[2mm]
&&\left\{
  \ba{l}
  D_\mu H=[\partial_\mu+A_\mu,H],\;H=\Phi+M,\\[2mm]
  Y=X+Y_0,\\[2mm]
  Y_0=H^2-M^2-\displaystyle{\sum_i}a_i[M^2,b_i].
  \ea
  \right.
\label{eqn:2-12}
\end{eqnarray}
See Ref. 7) for the most general gauge-invariant Lagrangian.
The potential term 
$
V={\rm tr}(1/g^2)Y^2
$
(except for the factor ${1\over 2}$)
is evaluated\cite{7)}
for the Glashow-Weinberg-Salam model
in the leptonic sector ($n_L=n_R=2$)
by assuming
the flavor
algebra
$\cA=C^\infty(M_4)\otimes(\bH\oplus\bC)$,
whose 
unitary group is $\cU(C^\infty(M_4)\otimes(\bH\oplus\bC))=$
Map($M_4,SU(2)\times U(1))$.
Writing
\begin{eqnarray}
&&\bA=\displaystyle{\sum_i}\rho(a_1^i,b_1^i)[D,\rho(a_2^i,b_2^i)],
\nn
\end{eqnarray}
we assume the following representation of \cA$\;$ on the chiral spinor
\begin{eqnarray}
&&\rho(a,b)=\left(
          \ba{cc}
          a&0\\
          0&B\\
          \ea
          \right)\otimes 1_{N_g},\;\;B=\left(
          \ba{cc}
          b&0\\
          0&b^{*}\\
          \ea
          \right),
\nn
\end{eqnarray}
where
$a=a(x)\in C^\infty(M_4)\otimes\bH,\;b=b(x)\in
C^\infty(M_4)\otimes\bC$
and
$*$ denotes the complex conjugation
so that the
left-handed and right-handed spinors belong to doublet 
$\psi_L=${\scriptsize$\left(
        \ba{l}
        \nu\\
        e\\
        \ea
        \right)_L$}
and singlet
$\psi_R=${\scriptsize$\left(
        \ba{l}
        \nu_R\\
        e_R\\
        \ea
        \right)$}, 
respectively, in $N_g$ generations.
Only doublets and singlets
appear in this model,
while the nonzero Abelian charge is quantized to be $\pm 1$.\cite{7)}
\footnote{
Similar quantization was also pointed out by
Hayakawa\cite{14)} in non-commutative QED.}
(In the present model $Y=0$ for $\psi_L$, $Y=+1$ for $\nu_R$,
and $Y=-1$ for $e_R$ provided that the hypercharge of Higgs doublet
is normalized to be $+1$
\footnote{
The correct hypercharge ($Y$) assignment of chiral leptons
will be discussed in the section 4
.}.)
Choosing the mass matrix as
\begin{eqnarray}
&&M=\left(
  \ba{cc}
  0&M_1\\
  M_1^{\dag}&0\\
  \ea
  \right),\;\;M_1=\left(
                  \ba{cc}
                  m_1&0\\
                  0&m_2\\
                  \ea
                  \right),\;\;
                  m_{1,2}:\;N_g\times N_g,
\nn
\end{eqnarray}
and putting
$1/g^2=${\scriptsize$
        \left(
        \ba{cc}
        (1/g_1^2)\otimes 1_{2N_g}&0\\
        0&(1/g_2^2)\otimes 1_{2N_g}\\
        \ea
        \right)$},
we find after making Connes' subtraction
or resorting to the auxiliary field method
\be
V=K(\phi^{\dag}\phi-1)^2,\;
K=\displaystyle{{1\over 2g_1^2}}{\rm 
tr}_g(m_1m_1^{\dag}+m_2m_2^{\dag})_\perp^2
+\displaystyle{{1\over g_2^2}}{\rm tr}_g[(m_1^{\dag}m_1)_\perp^2+
(m_2^{\dag}m_2)_\perp^2],
\label{eqn:2-13}
\ee
where
$
\phi=${\scriptsize$\left(
     \ba{cl}
     \phi_+\\
     \phi_0\\
     \ea
     \right)$}
is the normalized Higgs field and
tr$_g$ denotes the trace in the generation space with
tr$_gf_\perp^2=$tr$_gf^2-(1/N_g)($tr$_gf)^2$.
\\
\ind
We note that
$K=0$ for $N_g=1$ or $N_g>1$ with the
degenerate mass spectrum. For $N_g>1$ with
non-degenerate mass spectrum $K$ is positive.
We can take the vacuum expectation value
\footnote{By canonically
normalizing the Higgs kinetic energy term,
the vacuum expectation value of the
rescaled Higgs field
is proportional to the
quantity $\sqrt{{\rm tr}_g(m_1m_1^{\dag}+m_2m_2^{\dag})}$.
In the standard model this
implies that
the electroweak scale is essentially governed by the top mass,
which is not inconsistent with experiment.
The same remark will also apply for the potential
Eq.$\,$(\ref{eqn:2-18}) below.} 
of the
normalized Higgs field to be
$\langle\phi\rangle=${\scriptsize $
                                    \left(
                                    \ba{cl}
                                    0\\
                                    1\\
                                    \ea
                                    \right)$}.
\\
\ind
Next comes a generalization of
the wedge product (\ref{eqn:2-6})
of the Dirac matrices to include $\gamma^5=\gamma_5$.
A naive generalization
dismisses $\gamma^5$ when it appears twice
since $\gamma^5\wedge\gamma^5=(1/2)[\gamma^5,\gamma^5]=0$.
To avoid this we assign the `grade'
of $\gamma^A(A=\mu,5)$ by
$\epsilon_\mu=0 (\mu=0,1,2,3)$ and $\epsilon_5=1$ such that
the wedge product of $\gamma^5$ by itself
is converted into the anticommutator
$\gamma^5\wedge\gamma^5=(1/2)(\gamma^5\gamma^5-(-1)^{1\cdot 
1}\gamma^5\gamma^5)
=(1/2)\{\gamma^5,\gamma^5\}=1$.
This would give a sensible definition
of the field strength different from Connes' one.
\\
\ind
We found that the following definition works.
(Capital letters $A,B,C$ take the values 0,1,2,3,5.)
\begin{eqnarray}
f\wedge\gamma^A&\equiv& f\gamma^A=\gamma^Af=
\gamma^A\wedge f\;\;{\rm for}\;
{\rm any}\;{\rm complex}\;{\rm number}\;{\rm
or}\;{\rm function}\;f,\nn\\[2mm]
\gamma^A\wedge\gamma^B&=&
{\tilde A}_2[\gamma^A\gamma^B]\equiv
\displaystyle{1\over 2!}(\gamma^A\gamma^B
-(-1)^{\epsilon_A\cdot\epsilon_B}\gamma^B\gamma^A),\nn\\[2mm]
\gamma^A\wedge\gamma^B\wedge\gamma^C&=&
{\tilde A}_3[\gamma^A\gamma^B\gamma^C]\equiv
\displaystyle{1\over 3!}[\gamma^A\gamma^B\gamma^C
+(-1)^{\epsilon_A\cdot(\epsilon_B+\epsilon_C)}
\gamma^B\gamma^C\gamma^A\nn\\[2mm]
&&\qquad+(-1)^{\epsilon_C\cdot(\epsilon_A+\epsilon_B)}
\gamma^C\gamma^A\gamma^B
-(-1)^{\epsilon_A\cdot\epsilon_B}\gamma^B\gamma^A\gamma^C\nn\\[2mm]
&&\qquad
-(-1)^{\epsilon_A\cdot\epsilon_B+\epsilon_C\cdot(\epsilon_A+\epsilon_B)}
\gamma^C\gamma^B\gamma^A-(-1)^{\epsilon_C\cdot\epsilon_B}
\gamma^A\gamma^C\gamma^B],
\label{eqn:2-14}
\end{eqnarray}
where ${\tilde A}_n[\gamma^{A_1}
\gamma^{A_2}\cdots\gamma^{A_n}]$
denotes the graded antisymmetrization 
among the indices $A_1,A_2,$\\
$\cdots,
A_n$,
as indicated above for $n=2,3$. The number of the indices
$\mu=0,1,2,3$ is restricted to
less than or equal to
4.
We also define
\begin{eqnarray}
(\gamma^A\wedge\gamma^B)\wedge\gamma^C&=&
{\tilde A}_3[(\gamma^A\wedge\gamma^B)\gamma^C],\nn\\[2mm]
\gamma^A\wedge(\gamma^B\wedge\gamma^C)&=&
{\tilde A}_3[\gamma^A(\gamma^B\wedge\gamma^C)].
\label{eqn:2-15}
\end{eqnarray}
It is easy to check the associativity
\begin{eqnarray}
&&(\gamma^A\wedge\gamma^B)\wedge\gamma^C=
\gamma^A\wedge(\gamma^B\wedge\gamma^C)=
\gamma^A\wedge\gamma^B\wedge\gamma^C.
\label{eqn:2-16}
\end{eqnarray}
\ind
In terms of the wedge product of the Dirac matrices
defined above
we introduce the new field strength
\be
\bF=d\wedge\bA+\bA\wedge\bA,\;
d\wedge\bA\equiv\displaystyle{\sum_i}
[D,a_i]\wedge[D,b_i].
\label{eqn:2-17}
\ee
Connes' ambiguity problem still remains but,
since
$\bF=\bG|_{X=0}$,
we find the different result from
Eq.$\,$(\ref{eqn:2-13})
\be
V=
K'(\phi^{\dag}\phi-1)^2,\;\;\;
K'=\displaystyle{{1\over 2g_1^2}}{\rm tr}_g(m_1m_1^{\dag}+m_2m_2^{\dag})^2
+\displaystyle{{1\over g_2^2}}{\rm tr}_g[(m_1^{\dag}m_1)^2+
(m_2^{\dag}m_2)^2].
\label{eqn:2-18}
\ee
The Higgs potential is
generation-number independent
for
Eq.$\,$(\ref{eqn:2-17}).
\section{Extended differential formalism of Connes' YM on $M_4\times Z_2$}
\ind
One may inquire why there are two different 
field strengths for the non-commutative one-form.
In this section we shall give a non-commutative differential geometric
reason using
the extended differential formalism\cite{9),10)}
with
the extra one-form basis $\chi$.\cite{8)}
In this formalism,
in addition to the 
ordinary exterior derivative $d$
with $d\psi=\partial_\mu\psi d{\hat x}^\mu$, the hat
indicating the dimensionless coordinates,
we
define the extra exterior derivative $d_\chi$ by
\be
d_\chi\psi=M\psi\chi, 
\label{eqn:3-1}
\ee
where $M$ is the mass matrix in
Eq.(\ref{eqn:2-9}).
From the free Dirac Lagrangian
in the form
\begin{eqnarray}
&&\cL_D=i\langle{\tilde \psi},\bd\psi\rangle,\;\;\;\;
{\tilde \psi}=\gamma_\mu\psi d{\hat x}^\mu-\gamma_5\psi\chi,\;\;\;\;
\chi^{\dag}=-\chi,
\nn
\end{eqnarray}
where $\bd=d+d_\chi$ is the generalized exterior derivative
and $\langle d{\hat x}^\mu,d{\hat x}^\nu\rangle=\eta^{\mu\nu},
\langle d{\hat x}^\mu,\chi\rangle=\langle\chi,d{\hat x}^\mu\rangle=0,
\langle \chi,\chi\rangle=-1$,
we follow the prescription
in \S2
to obtain
\be
\cL_{\mbf D}=
i\displaystyle{\sum_i}
\langle (\rho^{\dag}(a_i){\tilde \psi}),
\bd(\rho(b_i)\psi)\rangle=i\langle \psi,
(\bd+\bA)\psi\rangle,
\label{eqn:3-2}
\ee
where
we have assumed
the Leibniz rule
\be
d_\chi(f\psi)=(d_\chi f)\psi+f(d_\chi\psi),
\;\;f=\rho(b_i),
\label{eqn:3-3}
\ee
used the condition (\ref{eqn:2-2}) and
defined the generalized gauge field $\bA$ by
\be
\bA=\displaystyle{\sum_i}a_i\bd b_i.
\label{eqn:3-4}
\ee
This is the non-commutative one-form
in the present notation.
In general, using the Leibniz rule (below) 
an arbitrary $n$-form\footnote{In this section
we use the usual notation
$\wedge$ for the exterior product.} is written
as
$\sum_ja_0^j\bd a_1^j\wedge \bd a_2^j\wedge\cdots\wedge \bd a_n^j$,
where $a_i^j (i=0,1,\cdots,n)$ are 0-forms similar to the function $f$.
\\
\ind
From Eqs.(\ref{eqn:3-1}) and (\ref{eqn:3-3})
we derive the action
of $d_\chi$ on $f$ 
\be
d_\chi f=[M,f]\chi,\;d_\chi(fh)=(d_\chi f)h+f(d_\chi h).
\label{eqn:3-5}
\ee
The antisymmetry
$
d{\hat x}^\mu\wedge d{\hat x}^\nu=-d{\hat x}^\nu\wedge d{\hat x}^\mu,\;
d{\hat x}^\mu\wedge \chi=-\chi\wedge d{\hat x}^\mu
$
ensures the nilpotency $d^2=0$ and
the relation
$
(dd_\chi+d_\chi d)f=(dd_\chi+d_\chi d)\psi=0$.
\\
\ind
There are two options to go further.
One is to
assume\cite{10)} the antisymmetry
also for the extra one-form basis
\be
\chi\wedge \chi=0.
\label{eqn:3-6}
\ee
The other instead assumes\cite{8),9)} the symmetry
\be
\chi\wedge \chi\not=0.
\label{eqn:3-7}
\ee
We now show that
these alternatives
lead to the field strength,
$\bG$ of Eq.(\ref{eqn:2-11}),
and the field strength, $\bF$ of Eq.(\ref{eqn:2-17}),
respectively.
\\
\ind
Let us first consider the symmetric case
(\ref{eqn:3-7}).
We define the action of the operator $\bd$
on the $n$-form through
$\bd(\sum_ja_0^j\bd a_1^j\wedge \bd a_2^j\wedge\cdots\wedge \bd a_n^j)=
\sum_j\bd a_0^j\wedge \bd a_1^j\wedge \bd a_2^j\wedge\cdots\wedge \bd 
a_n^j$.
Then $\bd$ is `nilpotent' in the sense that
$\bd(\bd a)=(\bd 1)\wedge(\bd a)=0$ because
$\bd 1=0$ due to the Leibniz rule.
However, this definition
leads to an 
ambiguity
$\bd(a_0\bd a_1)=\bd a_0\wedge\bd a_1\not=0$ even when
$a_0\bd a_1=0$.
\\
\ind
We next define\cite{9)}
the
field strength in this case
by the two-form
\be
\bF=\bd\wedge\bA+\bA\wedge\bA,\;
\bd\wedge\bA\equiv\displaystyle{\sum_i}\bd a_i\wedge\bd b_i,
\label{eqn:3-8}
\ee
which
turns out to be given by
\be
\bF=F+DH\wedge\chi+Y_0\chi\wedge\chi,
\label{eqn:3-9}
\ee
where $F=d\wedge A+A\wedge A$,
$DH=dH+[A,H]$
and $Y_0$ is given by Eq.(\ref{eqn:2-12}).
The bosonic Lagrangian
\begin{eqnarray}
&&\cL_B=-\langle\!\langle
\displaystyle{{1\over g^2}}\bF\,,\bF\,\,\rangle\!\rangle,
\;\;\;
\bF\,=F+DH\wedge\chi+Y_0\chi\wedge\chi
\nn
\end{eqnarray}
is evaluated by taking the inner product\cite{9)} of the 
two-form basis and performing the trace over the 2-dimensional
chiral space.
The result is the same as in Eq.$\,$(\ref{eqn:2-12})
with $Y\to Y_0$.
We recover
the generation-number independent
Higgs potential (\ref{eqn:2-18}) in this case.
\\
\ind
Next we consider the antisymmetric case (\ref{eqn:3-6}).
In this case the operator
$d_\chi$ is automatically nilpotent.
Thus the operator $\bd$ is also nilpotent, $\bd^2=0$,
so that
$\bd(\sum_ja_0^j\bd a_1^j\wedge \bd a_2^j\wedge\cdots\wedge \bd a_n^j)=
\sum_j \bd a_0^j\wedge \bd a_1^j\wedge \bd a_2^j\wedge\cdots\wedge\bd 
a_n^j$
holds true.
Consequently, $\bd(a_0\bd a_1)=0$ if $a_0\bd a_1=0$.
At first sight there seems to arise no ambiguity
problem encountered in the symmetric case.
However, the two-form field strength
now lacks the Higgs potential generating term at all.
\be
\bF=F+DH\wedge\chi.
\label{eqn:3-10}
\ee
We are then led to add a zero-form piece to the two-form
(\ref{eqn:3-10}) to define the field strength
by the Clifford product
\begin{eqnarray}
\bG&=&\bd\vee\bA+\bA\vee\bA=\bF+\bF_0,\nn\\[2mm]
\bF_0&=&\langle \bd,\bA\rangle+\langle \bA,\bA\rangle,\;
\langle \bd,\bA\rangle\equiv\displaystyle{\sum_i}\langle \bd a_i^{\dag},
\bd b_i\rangle.
\label{eqn:3-11}
\end{eqnarray}
This time the ambiguity problem
reappears because
$\langle \bd,\bA\rangle$ may not vanish
even when $\bA=0$.
The zero-form piece
$\bF_0=Y$ is given by Eq.(\ref{eqn:2-12}).
Using the fact
that two-form and zero-form are orthogonal,
the bosonic Lagrangian becomes
\begin{eqnarray}
\cL_B&=&-\langle\!\langle
\displaystyle{{1\over g^2}}\bG,\bG\,\rangle\!\rangle
=-\langle\!\langle
\displaystyle{{1\over g^2}}\bF,\bF\,\rangle\!\rangle
-V,\;\;V={\rm tr}\displaystyle{{1\over g^2}}Y^2,
\nn
\end{eqnarray}
where $\bF$ is defined by
Eq.(\ref{eqn:3-10}).
We thus obtain the same
result
as in the
previous
section
using Connes' field strength.
\section{Double sum prescription and the
standard model}
\ind
In this section we shall derive
Asquish's representation\cite{11)}
of Connes' color-flavor algebra of the standard model
\be
\cA=C^\infty(M_4)\otimes(\bH\oplus\bC\oplus M_3(\bC)),
\label{eqn:4-1}
\ee
whose unitary group is
$\cU(\cA)=$Map($M_4,U(3)\times SU(2)\times U(1))$,
from our formulation
using the double sum prescription.\cite{12)}
Here $M_3(\bC)$ denotes
the set of $3\times 3$ complex matrices.
\\
\ind
The algebra (\ref{eqn:4-1}) is represented on the doubled spinor
\be
\Psi=\left(
     \ba{l}
     \psi\\
     \psi^c\\
     \ea
     \right),\;\;\;\psi^c=C{\bar\psi},
\label{eqn:4-2}
\ee
where
$\psi$ stands for the total fermion field
\be
\psi=\left(
     \ba{l}
     \psi_L\\
     \psi_R\\
     \ea
     \right), \;\;
     \psi_L=\left(
     \ba{l}
     q_L\\
     l_L\\
     \ea
     \right), \;\;\psi_R=\left(
                         \ba{l}
                         u_R\\
                         d_R\\
                         \nu_R\\
                         e_R\\
     \ea
     \right).
\label{eqn:4-3}
\ee
We omit the color and generation indices for simplicity.
The free massive Dirac Lagrangian
is written in the present case ($n_L=n_R=8$)
as
\be
\cL_D=\displaystyle{{1\over 2}}
        {\overline{\it\Psi}}D{\it\Psi},\;\;
D=D_0+i\gamma_5\cM,\;\;
D_0=i\dslash\otimes 1_{32N_g},\;\;
\cM=\left(
      \ba{cc}
      M&0\\
      0&M^{*}\\
      \ea
      \right).
\label{eqn:4-4}
\ee
We choose the fermion mass matrix
as
\begin{eqnarray}
M_1&&=\left(
    \ba{cc}
    M_q\otimes 1_3&0\\
    0&M_l\\
    \ea
    \right),
    M_q=\left(
        \ba{cc}
        M_u&0\\
        0&M_d\\
        \ea
        \right),
     M_l=\left(
        \ba{cc}
        M_\nu&0\\
        0&M_e\\
        \ea
        \right).
\label{eqn:4-5}
\end{eqnarray}
This choice
is dictated by the global color symmetry and
the electric charge conservation. 
We assume Dirac mass $M_\nu$
for neutrinos.
\\
\ind
The product of the
$*$-preserving representations
$\rho_{1,2}$ is written as
\begin{eqnarray}
\rho(a,b,c)&=&\rho_1(a,b,c)\rho_2(a,b,c)=\rho_2(a,b,c)\rho_1(a,b,c),\nn\\[2
mm]
\rho_1(a,b,c)&=&\left(
              \ba{cc}
              \rho_w(a,b)&0\\
              0&\rho_s(b',c)\\
              \ea
              \right),\;\rho_w(a,b)=\left(
                \ba{cc}
                a\otimes 1_4&0\\
                0&B\otimes 1_4\\
                \ea
                \right)\otimes 1_{N_g},\nn\\[2mm]
\rho_2(a,b,c)&=&\left(
              \ba{cc}
              \rho_s^{*}(b',c)&0\\
              0&\rho_w^{*}(a,b)\\
              \ea
              \right),\;\;b'=b\;{\rm or}\;b^{*},
\label{eqn:4-6}
\end{eqnarray}
where
$(a, b, c)$ are the element of the 
algebra (\ref{eqn:4-1}) with
$c=c(x)\in C^\infty(M_4)\otimes M_3(\bC)$. 
The commutativity $\rho_1\rho_2=\rho_2\rho_1$
demands that $\rho_s(b',c)$ does not depend on
$a$.
Connes took\cite{3)} $b'=b$
for the case of massless neutrinos.
On the other hand, Asquish\cite{11)}
found for either massless or massive neutrinos
that the case $b'=b^{*}$
is also allowed from Poincar\'e duality.
We shall now derive Asquish's representation
and discuss implication of it
for the electric charge quantization.
\\
\ind
To this purpose
we generalize the prescription
in \S2
as follows.
To simplify the notation
let $a\in\cA$ such that
\begin{eqnarray}
\rho(a)&=&\rho_1(a)\rho_2(a)=\rho_2(a)\rho_1(a),\nn\\[2mm]
\rho_1(a)&=&\left(
                                \ba{cc}
                                \rho_w(a)&0\\
                                0&\rho_s(a)\\
                                \ea
                                \right),\;\;
\rho_2(a)=\left(
                                \ba{cc}
                                \rho_s^{*}(a)&0\\
                                0&\rho_w^{*}(a)\\
                                \ea
                                \right).
\label{eqn:4-7}
\end{eqnarray}
To make Eq.$\,$(\ref{eqn:4-4}) gauge-invariant under
the gauge transformation
\be
     {\it \Psi}\to \rho(g){\it \Psi}=\rho_1(g)\rho_2(g){\it \Psi},\;\;\;
     {\overline{\it \Psi}}\to{\overline{\it \Psi}}\rho^{\dag}(g)
     ={\overline{\it 
\Psi}}\rho_2^{\dag}(g)\rho_1^{\dag}(g),\;\;\;g\in\cU(\cA),
\label{eqn:4-8}
\ee
we 
consider the non-symmetry transformations
\be
     {\it \Psi}\to\rho_1(b_i)\rho_2(b_j){\it\Psi},\;\;\;
     {\overline{\it\Psi}}\to{\overline{\it\Psi}}
     \rho_1(a_i)\rho_2(a_j),\;\;\;a_i,a_j,b_i,b_j\in\cA
\label{eqn:4-9}
\ee
and take the double sum\cite{12)} over the indices $i$ and $j$ 
after substituting Eq.$\,$(\ref{eqn:4-9}) into the free massive
Dirac Lagrangian (\ref{eqn:4-4}) to maintain
the equal-time anticommutation relations
by the condition
\be
\displaystyle{\sum_{i,j}}\rho_1(a_i)\rho_2(a_j)
\rho_1(b_i)\rho_2(b_j)=\big(\displaystyle{\sum_i}\rho_1(a_i)\rho_1(b_i)\big
)
\big(\displaystyle{\sum_j}\rho_2(a_j)\rho_2(b_j)\big)=1.
\label{eqn:4-10}
\ee
Take, for instance, $a_1^{\dag}=b_1=g_1\in \cU(\cA)$
and $a_{i\not=1}=b_{i\not=1}=0$ inside the first round bracket.
Then the first factor equals unity, implying the second factor
to be equal to 1. Consequently, we have the general conditions
\be
\displaystyle{\sum_i}\rho_1(a_i)\rho_1(b_i)=
\displaystyle{\sum_j}\rho_2(a_j)\rho_2(b_j)=1.
\label{eqn:4-11}
\ee
We then get the result
\begin{eqnarray}
\cL_{\mbf D}&=&\displaystyle{{1\over 2}}
{\overline{{\it \Psi}}}(D_0+i\gamma_5\cM+\bA){\it \Psi},
\bA=\displaystyle{\sum_{i,j}}\rho_1(a_i)\rho_2(a_j)
[D,\rho_1(b_i)\rho_2(b_j)]\equiv\tbA+\tbA^c,
\nn\\[2mm]
\tbA&=&\displaystyle{\sum_i}\rho_1(a_i)
[D,\rho_1(b_i)],\;\;\;
     \tbA^c=\displaystyle{\sum_j}\rho_2(a_j)
[D,\rho_2(b_j)],
\label{eqn:4-12}
\end{eqnarray}
where we have assumed that
\be
[[\cM,\rho_1(b_i)],\rho_2(b_j)]=0.
\label{eqn:4-13}
\ee
It turns out that
this is equivalent to the
condition\cite{11)}
\footnote{The condition (\ref{eqn:4-13}) is assumed only for Dirac mass 
terms.
Majorana mass terms for neutrinos do not obey
this condition and
lead to Higgs triplet and singlet.}
from Poincar\'e duality.
Putting
\begin{eqnarray}
A&=&\displaystyle{\sum_i}\rho_w(a_i)[D_0,\rho_w(b_i)],\;\;\;
\Phi=\displaystyle{\sum_i}\rho_w(a_i)[M,\rho_w(b_i)],\nn\\
A^c&=&\displaystyle{\sum_j}\rho_s(a_j)[D_0,\rho_s(b_j)],\;\;\;
\Phi^c=\displaystyle{\sum_j}\rho_s(a_j)[M,\rho_s(b_j)],
\label{eqn:4-14}
\end{eqnarray}
the gauge-invariant Dirac Lagrangian (\ref{eqn:4-12}) becomes 
with $A^{c*}\equiv i\gamma^\mu A_\mu^{c*}$
\be
\cL_{\mbf D}={\bar\psi}(D+A+A^{c*}+i\gamma_5({\it \Phi}
+{\it \Phi}^{c*}))\psi,\;\;
D=D_0+i\gamma_5 M,\;\;D_0=i\dslash\otimes 1_{{\rm dim}\rho_w}.
\label{eqn:4-15}
\ee
\ind
We are now in a position to determine the representation
$\rho_s(b',c)$
based on Eq.$\,$(\ref{eqn:4-13})
which means
\be
[[M,\rho_w(a,b)],\rho_s^{*}(b',c)]=0.
\label{eqn:4-16}
\ee
The reasoning is the same as in Ref. 11).
Using the mass matrix (\ref{eqn:4-5})
and writing
\begin{eqnarray}
\rho_w(a,b)&=&\left(
                \ba{cc}
                \rho_{wL}(a)&0\\
                0&\rho_{wR}(b)\\
                \ea
                \right)\otimes 1_{N_g},\nn\\[2mm]
\rho_s^{*}(b',c)&=&\left(
                \ba{cc}
                \rho_{sL}(b',c)&0\\
                0&\rho_{sR}(b',c)\\
                \ea
                \right)\otimes 1_{N_g},
\nn
\end{eqnarray}
we obtain the equation
\begin{eqnarray}
&&(M_1\rho_{wR}-\rho_{wL}M_1)\rho_{sR}-\rho_{sL}(M_1\rho_{wR}-\rho_{wL}M_1)
=0.
\nn
\end{eqnarray}
(It is enough to consider the case $N_g=1$.\cite{11)})
Since only $\rho_{wL}$ depends on $a$ this is equivalent to
two conditions
\begin{eqnarray}
\rho_{wL}M_1\rho_{sR}-\rho_{sL}\rho_{wL}M_1&=&0,\nn\\[2mm]
M_1\rho_{wR}\rho_{sR}-\rho_{sL}M_1\rho_{wR}&=&0.
\label{eqn:4-17}
\end{eqnarray}
The commutativity $\rho_1\rho_2=\rho_2\rho_1$
implies $[\rho_{wL},\rho_{sL}]=[\rho_{wR},\rho_{sR}]=0$
so that Eq.$\,$(\ref{eqn:4-17})
is reduced to
a single equation
\begin{eqnarray}
M_1\rho_{sR}-\rho_{sL}M_1=0,
\nn
\end{eqnarray}
from which we deduce that
\begin{eqnarray}
\rho_{sL}(b',c)&=&\rho_{sR}(b',c)=\left(
\ba{cc}
1_2\otimes c&0\\
0&b'^{*}1_2\\
\ea
\right)\otimes 1_{N_g}.
\nn
\end{eqnarray}
Consequently, we obtain\cite{11)}
\be
\rho_s(b',c)=\left(
                \ba{cccc}
                1_2\otimes c^{*}& & &\bigzerol\\
                &b'1_2 &&  \\
                & & 1_2\otimes c^{*}&\\
                \bigzeroL& & &b'1_2\\
                \ea
                \right)\otimes 1_{N_g},\;\;\;b'=b\;{\rm or}\;b^{*}.
\label{eqn:4-18}
\ee
This implies that $\Phi^c=0$
in Eq.$\,$(\ref{eqn:4-14}) so that
the strong force associated with $\rho_s(b',c)$
is vectorial.\cite{11)}
In other words,
the gauge-invariant Dirac Lagrangian (\ref{eqn:4-15}) becomes 
\be
\cL_{\mbf D}={\bar\psi}(D_0+A+A^{c*}+i\gamma_5H)\psi,\;\;\;
H=M+{\it \Phi}.
\label{eqn:4-19}
\ee
It can be shown that this is the well-known
standard model Dirac Lagrangian.
We do not feel it necessary to discuss the bosonic sector
any more. Rather we focus upon the charge quantization
problem in the light of Eq.$\,$(\ref{eqn:4-18}).
\\
\ind
The only ambiguity in our derivation
of Eq.$\,$(\ref{eqn:4-19})
is the appearance of $b'$ in the
representation (\ref{eqn:4-18}).
We shall now show that
the correct hypercharge assignment
is obtained irrespective of the choice $b'=b$ or $b^{*}$.
The gauge transformation
\be
\psi\to ^g\psi=\rho_w(a,b)\rho_s^{*}(b',c)\psi,
\label{eqn:4-20}
\ee
where
$(a,b,c)$ is the element of $\cU(\cA)$, namely,
$a=u\in SU(2), b=e^{i\alpha}, c=e^{i\beta}v, v\in SU(3)$,
$\alpha$ and $\beta$ being real,
contains two $U(1)$ factors.
If 
the gauge transformation (\ref{eqn:4-20}) is unimodular
\be
{\rm det}\rho_w(a,b)\rho_s^{*}(b',c)=1,\;(a,b,c)\in\cU(\cA),
\label{eqn:4-21}
\ee
the $\alpha$ and $\beta$ are related such that only
$U(1)_Y$
survives with tr$Y=0$ per generation.
The unimodularity condition (\ref{eqn:4-21})
implies det$\rho_s(b',c)=1$ for unitary $(b',c)$,
leading to $3\beta\mp\alpha=0$, the sign
depending on $b'=b$ or $b^{*}$.
The minus sign ($b'=b$) implies
the usual assignment of the hypercharge
(the coefficient of $\alpha$
\footnote{
The hypercharge of Higgs doublet is normalized to be +1.}),
$Y(l_L)=-1, Y(\nu_R)=0, Y(e_R)=-2, Y(q_L)=1/3, Y(u_R)=4/3$ and 
$Y(d_R)=-2/3$.
On the other hand,
the plus  sign ($b'=b^{*}$)
gives a different set of values,
$Y(l_L)=+1, Y(\nu_R)=+2, Y(e_R)=0,
Y(q_L)=-1/3, Y(u_R)=2/3$ and $Y(d_R)=-4/3$.
It can be shown, however, that
the renaming
$l_L=${\scriptsize{$\left(
                              \ba{l}
                              \nu\\
                              e\\
                              \ea
                              \right)_{{\mbox{\scriptsize}}L}$}} 
$\to l_R^c=${\scriptsize{$\left(
                              \ba{l}
                              e^c\\
                              -\nu^c\\
                              \ea
                              \right)_{{\mbox{\scriptsize}}R}$}}
and $l_R=${\scriptsize{$\left(
                              \ba{l}
                              \nu_R\\
                              e_R\\
                              \ea
                              \right)$}} 
$\to l_L^c=${\scriptsize{$\left(
                              \ba{l}
                              e^c_L\\
                              -\nu^c_L\\
                              \ea
                              \right)$}}
together with $M_\nu\leftrightarrow M_e^{*}$ and $\rho_s\to\rho_s^{*}$
and similarly for quarks converts the second solution 
$Q(\nu)=1, Q(e)=0, Q(u)=1/3, Q(d)=-2/3$ to the first one
$Q(e^c)=1,Q(\nu^c)=0, Q(d^c)=1/3, Q(u^c)=-2/3$,
where $Q(f)$ denotes the electric charge of the fermion $f$.
That is, the electric charge quantization
is linked to the single unimodularity condition (\ref{eqn:4-21})
for the case of massive neutrinos
\footnote{
This conclusion solely depends on Asquish's representation
and remains true even if our double sum prescription turns out to be 
wrong.}.
\\
\ind
In contrast, the case is not true for massless
neutrinos.
In fact, we should replace Eqs.$\,$(\ref{eqn:4-6})
and (\ref{eqn:4-18})
with 
\begin{eqnarray}
\rho_w(a,b)&=&\left(
                \ba{ccc}
                a\otimes 1_4&0&0\\
                0&B\otimes 1_3&0\\
                0&0&b^{*}\\
                \ea
                \right)\otimes 1_{N_g},\nn\\[2mm]
\rho_s(b,c)&=&\left(
                \ba{cccc}
                1_2\otimes c^{*}& & &\bigzerol\\
                &b'1_2 &&  \\
                & & 1_2\otimes c^{*}&\\
                \bigzeroL& & &b'\\
                \ea
                \right)\otimes 1_{N_g},\;\;\;b'=b\;{\rm or}\;b^{*}.
\label{eqn:4-22}
\end{eqnarray}
The unimodularity condition (\ref{eqn:4-21})
then leads to $3\beta-\alpha=0$
for $b'=b$ but to $6\beta+\alpha=0$ for $b'=b^{*}$.
The case $b'=b$ implies
the usual assignment of the hypercharge.
However, the case $b'=b^{*}$
gives the anomaly-non-free solution
$Y(l_L)=+1, Y(e_R)=0, Y(q_L)=-1/6, Y(u_R)=5/6$ and $Y(d_R)=-7/6$.
To summarize we have found that
the single unimodularity condition
(\ref{eqn:4-21}) leads to
the anomaly-free solution
provided that $\nu_R$ exist in each generation.
\section{Discussion}
\ind
The present paper concerned with
a field-theoretic prescription for
Connes' YM on $M_4\times Z_2$ which {\it derives Higgs
from the Dirac operator
but does not assume Higgs as an input element of the theory}.
Our reformulation based on
the local non-symmetry transformations
greatly simplifies Connes' mathematical presentation
and achieves the
unification of the gauge
and Higgs fields without the axioms of NCG.
\\
\ind
Incidentally, we also found that
the field strength
in Connes'
gauge theory is not unique
and there are two definitions possible.
Connes' definition
leads to the
generation-number dependent
Higgs potential,
while our definition
yields the
generation-number independent
Higgs potential.
\\
\ind
It can be shown that 
in the standard model only Higgs doublet, triplet and singlet
are allowed in our formulation
because our method generates only Higgs coupled to chiral fermions.
(Higgs triplet and singlet
can appear only for massive neutrinos with Majorana masses.)
It is an open question whether or not
our method is generalizable to
describe GUT which contains Higgs without Yukawa coupling to chiral 
fermions.
We postpone this problem to a future work.
\section*{Acknowledgements}
One (K.M.) of the authors is grateful to
Professor S. Kitakado
for useful discussions and
continuous encouragement.

\end{document}